# Single dot photoluminescence excitation spectroscopy in the telecommunication spectral range

Paweł Podemski[a,*], Aleksander Maryński[a], Paweł Wyborski[a], Artem Bercha[b], Witold Trzeciakowski[b], Grzegorz Sęk[a]

[a] *Laboratory for Optical Spectroscopy of Nanostructures, Department of Experimental Physics, Faculty of Fundamental Problems of Technology, Wrocław University of Science and Technology, Wybrzeże Wyspiańskiego 27, 50-370 Wrocław, Poland*

[b] *Institute of High Pressure Physics, Polish Academy of Sciences, Sokołowska 29/37, 01-142 Warsaw, Poland*

**Abstract**

Single dot photoluminescence excitation spectroscopy provides an insight into energy structure of individual quantum dots, energy transfer processes within and between the dots and their surroundings. The access to single dot energy structure is vital for further development of telecom-based quantum emitters, like single photon sources or entangled pair of photons. However, application of single dot photoluminescence excitation spectroscopy is limited mainly to dots emitting below 1 μm, while nanostructures optically active in the telecommunication windows of 1.3 and 1.55 μm are of particular interest, as they correspond to the desirable wavelengths in nanophotonic applications. This report presents an approach to photoluminescence excitation spectroscopy covering this application-relevant spectral range on single dot level. Experimental details are discussed, including issues related to the tunable excitation source and its spectral filtering, and illustrated with examples of photoluminescence excitation spectroscopy results from single quantum dots emitting in both the 1.3 and 1.55 μm spectral ranges.

---

* Corresponding author.
*E-mail address:* pawel.podemski@pwr.edu.pl (P. Podemski).

# 1. Introduction

Since the first proposal by Arakawa and Sakaki in 1982 [1] quantum dots (QDs) have passed a long way from purely theoretical concept to real world applications. The tremendous development of epitaxial techniques and processing of semiconductor structures allowed QDs to find application in many areas of modern optoelectronic solutions. There have been already demonstrated QD-based lasers operating in O band (1.3 µm) [2] and C band (1.55 µm) [3] or broadband semiconductor optical amplifiers [4]. All these devices rely on the averaged properties of many QDs, by and large in a few stacked layers. However, the unique attributes of QDs are even more manifested when the device operation principle is based on a single dot. There has been a great effort put in the development of telecom-wavelength-compatible single photon sources [5-9] – the only way to assure the security of the quantum cryptography protocols and an important part of realization of quantum information processing devices for their utilization in the quantum networks based on the existing optical fiber connections. However, for further development there are still needed efforts in engineering of QDs, including their morphology, surface density, electronic structure and emission purity.

The fabrication of nanophotonic devices requires a precise information on individual quantum dot properties, especially on their confined states structure, directly related, i.e., to cascades of exciton complexes resulting in single or entangled photons generation. The knowledge on the QD excited states is also important as they are directly related to temperature stability of the devices. Furthermore, some of the properties of quantum sources (e.g. single photon emission purity, photons indistinguishability or spectral linewidth) can be improved by quasi-resonant excitation, where the knowledge on the exact QD excited state energy is crucial. To acquire this information it is necessary to take advantage of an experimental technique capable of probing the confined states of a single dot. Typically, it would be an absorption measurement, but unfortunately absorption of an individual QD is very small and the dots are usually embedded in a complex and not so transparent structure. Nevertheless, there are absorption-like techniques, which can overcome the problem with low absorption of single QDs, like photoluminescence excitation spectroscopy (PLE). This technique is based on an emission (photoluminescence) measurement, where the excitation energy is tuned, probing different states in the energy spectrum of the investigated structure. The resulting spectrum is a product of four factors: intensity of the exciting source, probability of the absorption at the given energy, probability of the transfer of optically generated carriers to the emitting state and probability of the radiative recombination at this state. The first and the last factors directly influence a typical photoluminescence spectrum, while the remaining two are not varying during this kind of an experiment. In the PLE measurement, however, the excitation energy is tuned and also these two middle factors do change. In this way it is possible to track the absorption dependence on the energy and/or follow the carrier transfer between different parts (and energy states) of the examined structure, what is directly related to the quantum efficiency or the generation rate of quantum sources.

For single QD study, PLE has been used for a long time, identifying the energy structure of confined levels [10–14], demonstrating phonon-enhanced carrier relaxation [15], showing existence of continuum of states at the energies below the wetting layer ground state where only sharp absorption lines are usually expected [16], evidencing coupling between single QDs [17–19] or making possible coherent control of a single quantum dot [20]. However, these studies have been limited to dots emitting below 1 μm, primarily. The main limitation of the optical access to single dots emitting at longer wavelengths (and especially at the telecom range) is the tuning range of Ti:sapphire laser (approximately 700-1000 nm), which is the most common tool for probing the single dot properties. Beyond this range, there is a necessity of an appropriate tunable excitation source, followed by other changes needed in the experimental setup to accommodate to a new and more experimentally demanding spectral range. So far, there was no approach proposed and described in detail to extend single dot PLE to longer wavelengths. One of the solutions is to use an optical parametric oscillator pumped by a Ti:sapphire laser [6], nonetheless the pulsed nature of this excitation results in a rather large laser linewidth, providing low selectivity while exciting different states within QD, and less efficient excitation due to the small filling factor leading to low average power, despite high power in a single pulse. For single dot continuous wave PLE above 1 μm, there is only an attempt on InAs/InP quantum dot within a horn structure [21,22], however with no information on how the single dot PLE could be realized in this spectral range. This kind of experiment above 1 μm requires a more demanding experimental setup construction and additional factors which have to be taken into account, before a PLE spectrum from a single dot can be recorded. This report attempts to fill this gap and presents the details of a single dot PLE experiment in the spectral range above 1 μm, including the realization of single dot tunable excitation and its spectral filtering. As an illustration of efficiency of this approach, there are presented examples of PLE results from single QDs emitting in the telecom O and C bands.

## 2. Experiment

2.1. Tunable excitation laser source

One of the most challenging parts of single dot PLE beyond Ti:sapphire laser excitation spectral range, is to provide excitation source which is continuous wave, tunable with high accuracy and offering narrow linewidth and sufficient output power. Our solution is based on semiconductor laser diodes matching the desired excitation spectral range, which are employed in an external cavity system. This design provides continuous tuning of the excitation wavelength in the range of the order of tens of nanometers, which is usually sufficient for single QD properties study. In general, there are two main approaches for a tunable laser with an external cavity. In Littrow configuration, optical cavity is composed from back cleaved edge of the laser diode and a rotating diffraction grating as the end mirror [23]. The laser diode receives optical feedback from the first-order diffracted from the grating and the tuning is realized by rotation of the grating. However, this arrangement while relatively simple, changes the direction of the output beam while the laser is tuned, which is a serious problem during PLE measurement, where individual QDs have to be addressed with a laser spot of single micrometers size. It can be overcome by adding an additional mirror at the laser exit, which actively accommodates for changes of the beam angle. Another approach is the Littman configuration, where the diffraction grating is fixed and an additional rotating mirror is added, which reflects the first-order beam back to the laser diode [24,25]. This solution provides stable output beam since the wavelength is tuned by mirror rotation only, nevertheless, the grating is passed twice resulting in lower output power, compared to the Littrow configuration. For single dot PLE study in the telecom range, we tested both the configurations. For the C band emitting QDs there was used a commercial external cavity laser in Littrow configuration with the tuning range of 1440-1540 nm with output power up to 80 mW and the full width at half maximum (FWHM) below 1 nm. The excitation line for one of the wavelength values from the tuning range is shown in Fig. 1. (a). The output beam stability was provided by automated mirror following the diffraction grating angle change. QDs for the O band were excited by self-made external cavity laser in the Littman arrangement based on a curved stripe gain chip and a diffraction grating with 900 grooves/mm and 500 nm blaze, providing 50% of first-order reflection for 1.3 μm wavelength. This solution provided tuning range of 1210-1310 nm, output power up to 70 mW and the FWHM below 1 nm (see Fig. 1. (b)), with the possibility of further improvements of output parameters by an interplay between the tuning range, output power and the laser linewidth by realignment of the external cavity or exchange of the diffraction grating. The self-made construction significantly reduces the cost of an excitation source, providing at the same time an open system for custom arrangements, e.g. laser diode exchange to easily adapt to the required tuning spectral range.

The lasing output from both these types of external cavity lasers is accompanied by a low-level background (typically 40 dB lower than the peak intensity), which is negligible in most spectroscopic measurements. Nonetheless, when single dot emission is investigated this background becomes a serious hindrance. Figure 2 shows an example of QD emission spectra with excitation by an external cavity laser. As a reference, there is also shown a single dot spectrum obtained under nonresonant (639 nm) excitation by a standard diode laser – the bottommost spectrum. During the measurement there was used a longpass filter in the detection part of the setup to filter out the laser line before reaching the detector (the edge of the filter is marked in Fig. 2 by the dashed line). When the excitation is changed to external cavity laser (middle spectrum), the QD emission becomes overlapped by the background from the tunable laser, what makes the PLE measurement and analysis very difficult. In Fig. 2, there is only a slope of the laser background visible (it was scaled down by a factor of 30), which dominates the whole QD spectrum. Moreover, its intensity and shape depend on the specific wavelength, resulting in practically unusable PLE spectra. To overcome this issue, we have employed a laser line filtering with short-focal-length monochromator (300 mm) placed next to the external cavity tunable laser combined with shortpass filters, providing initial background removal. The topmost spectrum in Fig. 2 shows a single QD emission after the exciting laser line was filtered by a short-focal-length monochromator. The emission spectrum shows similar picture to the nonresonant (639 nm) excitation with some selectivity of the emission lines, as expected for the resonant excitation (e.g. in a QD excited state), and without any trace of the laser background. The excitation power was increased to 100 μW, compared to the nonresonant case (bulk absorption), to compensate for lower absorption by the quantum dots and slightly larger laser spot on the sample surface.

To align the tunable laser and monochromator we have verified a few different approaches. The most accurate wavelength measurement (of the order of single picometers), which is then sent to the monochromator, is provided by wavelength meters, where the measurement is based on Fizeau interferometers. This solution requires, however, pure single-mode laser operation, which is not always a case in the whole tuning range. The mutual tuning of a laser and the filtering monochromator does not require such high precision in wavelength determination, so we have also used a fixed-grating infrared spectrometer, which was sufficient as a reference for the laser line filtering. When the backlash of the external cavity laser mechanical tuning can be neglected during the measurement, there is no necessity to provide live laser wavelength measurement and the filtering monochromator can be calibrated to the excitation source in advance. The typical power loss during the filtering process is of the order of 25%, which still provides available excitation power up to 20 mW, more than sufficient for single dot study, and can even be further optimized if needed, with brighter monochromator diffraction gratings or shorter monochromator focal length. The spectral filtering can be also realized by coupling to a single-mode optical fiber, however our tests showed the efficiency of this solution is strongly dependent on the alignment of the whole experimental setup and is in most cases less effective than the monochromator-based filtering.

2.2. Experimental setup

The tunable laser was introduced into a high spatial resolution photoluminescence (i.e. microphotoluminescence) setup, adapted to the varying excitation wavelength. The scheme of the whole setup is shown in Fig. 3. To minimize the chromatic aberration, the optical path was based on mirrors rather than lenses. Only in a few spots, where it was difficult to use mirrors, achromatic optics was used instead. A few percent of the chosen tunable laser (1 or 6) was directed to the fixed-grating infrared spectrometer or the wavelength meter (9), providing reference wavelength for the filtering monochromator (12) calibration. The remainder of the laser beam was optionally sent through shortpass filter (10), which can decrease the unwanted laser background but its availability is strongly wavelength-dependent, and filtered by the short-focal-length (300 mm) monochromator (12). At this point a fixed-wavelength laser (e.g. 639 nm) (13) could be introduced into the optical path, replacing the tunable laser and providing standard microphotoluminescence measurement as a reference to PLE spectra. The laser power was controlled by a motorized neutral density gradient filter (14) coupled to the power meter head (15) and guided to a microscope objective (17) by a spectrally-matched non-polarizing beamsplitter cube (16). The microscope objective (17) was achromatic (480-1800 nm), infinity corrected, with working distance of 2 cm and 0.4 numerical aperture, resulting in a laser spot diameter on the sample surface of single micrometers. It was mounted on a stage with manual and piezo-controlled movement, providing spatial resolution of the order of 10 nm. The sample was mounted in a continuous-flow microscopy cryostat (18) with short working distance and providing temperatures down to 5 K. The cryostat was put on a motorized stage with the movement precision below 100 nm, which together with piezo-controlled microscope objective provided accurate alignment, required for single dot study. Emission from a sample was collected by the same microscope objective (17), sent through a longpass filter (22) removing the laser line, and focused by an achromatic doublet lens (23) on an entrance slit of a monochromator (24) with focal length of 1000 mm. The monochromator was coupled to nitrogen-cooled linear array detector of 1024 InGaAs pixels (25), offering spectral resolution below 100 μeV. Alternatively, the optical path can be redirected to the infrared imaging camera (21) for high-magnification view of the sample surface, which provides a way of a direct selection of the excitation/collection spot on the sample surface. In this mode, infrared LED (19), introduced through a beamsplitter with a few percent of reflection (8), acts as a light source illuminating surface of the sample using Köhler illumination method [26].

A photoluminescence excitation spectrum raises from the consecutive emission measurements, while the excitation wavelength is altered. However, this wavelength change is not straightforward when single QDs are addressed. Since the excitation laser needs to be filtered, the short-focal-length monochromator has to be adjusted properly for every new wavelength, what requires former acquisition of the new wavelength value by the fixed-grating infrared spectrometer or the wavelength meter. Additionally, the wavelength tuning moves over the laser gain, resulting in the varying laser power. In PLE experiments, this is sometimes overcome by a measurement of the excitation power wavelength dependence and the PLE spectra are normalized with regards to the changing power after the measurement is performed. Nevertheless, in a case of a more sensitive experiment (e.g. involving single QDs) the power has to be normalized during the measurement process, to avoid the excessive excitation of the investigated structures and to assure constant excitation power independently of the wavelength. The change of excitation power results in an effectively different excitation area on the sample surface, leading to the observation of additional QDs (located further from the excitation spot center) for higher power values. Additionally, in QDs the relaxation between states is significantly slower than in bulk materials, or even in quantum wells, due to the phonon bottleneck effect [27]. The uncontrolled increase in the excitation power can result in the qualitative change of the PLE spectra induced by the blocked relaxation channels from the states that are excited. Finally, the excessive excitation power can even lead to the effects related to the excitation-dependent response saturation, i.e. resulting in the nonlinear response regime. These issues cannot be removed in the post-processing of the recorded data by simple spectra renormalization. Hence, every change of the excitation wavelength needs to be followed by a few adjustment steps, which extend significantly the measurement time when multiplied by the number of the excitation wavelength values. Despite the simple inconvenience of such an experiment it can also influence the results directly. When the measurement time becomes excessively long, there are becoming more important effects of sample temperature stability or spatial drift (moving QDs away from the observed spot). To overcome these problems, we have developed software to fully automatize the measurement. In the PLE measurement process, firstly the excitation laser wavelength is changed, then a new wavelength is detected by the fixed-grating infrared spectrometer or the wavelength meter and sent to the short-focal-length monochromator, responsible for the laser line filtering. Next, the power is corrected to the requested value and the measurement by the InGaAs linear array is triggered. The procedure is repeated for every new excitation wavelength, decreasing significantly the measurement time and providing more consistent results.

2.3. Examples of single dot photoluminescence excitation spectra in the telecommunication spectral range

To illustrate the single dot PLE measurement, two QD samples were used. For O band (1.3 μm), these were $In_{0.8}Ga_{0.2}As$ quantum dots grown by MOCVD on a GaAs substrate and capped with an $In_{0.2}Ga_{0.8}As$ strain-reducing layer (with nominal thickness of 4 nm) [6]. Resulting density of QDs was of the order of $10^8$ cm$^{-2}$ with the dots base diameter of 30 nm and the average height of 6-8 nm. As quantum dots emitting in C band (1.55 μm), there were used InAs QDs grown by MBE on InP substrate [28]. The density of these QDs was $2 \cdot 10^9$ cm$^{-2}$ with relatively large size variation (typical height of 30 nm). All the measurements were performed at 5 K.

Figure 4 shows single dot PLE results on InGaAs/GaAs QDs emitting in 1.3 μm spectral region with the excitation power of 200 μW (measured outside of the cryostat). On the PLE map (Fig.4. (a)) there can be distinguished single dot emission lines with a few clear resonances. These lines arise from ground states of different quantum dots (varying slightly in size, resulting in the shifted emission energy) and, possibly, different exciton complexes within individual QDs. The intensity change of one of the emission lines (0.9258 eV) is presented in Fig. 4. (b), with a pronounced intensity increase at the excitation energy of about 75 meV higher than the emitting state. It can be explained when confronted with energy structure calculations for such quantum dots [29], where the single particle states were calculated within 8-band **k·p** model with realistic structure parameters (size and inhomogeneous indium composition), taking into account the influence of the strain within the continuous elasticity approach and the second order of the piezoelectric effect, with the excitonic structure obtained in the configuration-interaction method, providing energy difference between QD first excited (p-shell) and ground (s-shell) states. The calculations indicate on the excited state origin of this feature – when QD is resonantly excited in the p-shell state, the generated carriers efficiently feed the ground state, from which the emission is being observed. This observation is also in agreement with the power-dependent photoluminescence spectra from the ensemble of this kind of quantum dots, where the first excited state was observed at the energy of 67 meV above the QD ground state – see ref. [29]. The linewidths of the observed PLE resonances are relatively broad (2 meV for the maximum in Fig. 4. (b)). It most likely results from the excitation laser linewidth (about 1 nm – see Fig. 1. (b)), but can be also related to the more complex energy structure of the excited states spectrum [29], leading to a broader PLE response. The detailed study of the energy structure of these QDs will be reported elsewhere.

In Fig. 5, there are shown PLE results for single InAs/InP QDs emitting at longer wavelengths – close to 1.55 μm. The excitation power was of the similar order (500 μW) and, likewise, in Fig. 5. (a) there are observed single dot emission lines. The PLE spectrum for one of the emission lines (0.7977 eV) is presented in Fig. 5. (b). This kind of spectra were recorded also for other QDs in this structure with two characteristic features observed. There is, again, strong intensity increase at the excitation energy 20-50 meV higher than the emitting state (35 meV in the presented case), which is probably related to the p-shell absorption (however, we do not have energy structure simulation data for this kind of quantum dots). The shown maximum is composed of a couple of peaks, which is most likely related to the presence of a few different exciton complexes in both ground (emitting) and p-shell (being excited) states, what could be resolved due to the narrow enough excitation laser line. There is also a weaker band (26-32 meV in the shown spectrum), which keeps constant energy difference to the exciting laser and follows it energetically, while the laser is being tuned. This can indicate a phonon-assisted effect, where the absorption is at the virtual state, higher than the emitting one by the optical phonon energy, and then the carriers relax efficiently (by emitting phonons) to the ground state, observed in this PLE experiment. Nonetheless, the observed band is surprisingly broad and demonstrates some substructure. However, all details of these effects are beyond the scope of this paper and will be examined independently.

## 3. Summary and outlook

There was presented an approach to photoluminescence excitation spectroscopy on single dot level in the spectral range of 1.3 and 1.55 μm telecommunication windows. It opens a way for investigation of single nanostructures in this application relevant spectral range (e.g. for optical-fiber-compatible single photon sources), providing information on their confined states structure and energy transfer processes. There were discussed details of the experiment, illustrated with exemplary results of single dot PLE in both 1.3 and 1.55 μm spectral regions, showing the capabilities of the single dot PLE setup.

Still, there is a room for further improvements to increase the quality of the spectra and to enhance the spectral range, which can be investigated. For instance, the demonstrated excitation laser linewidth (especially in Littman configuration) is still relatively broad – hundreds of microelectronvolts is a few times more than a typical single QD emission linewidth. It becomes a serious hindrance while probing energy states that are spectrally very close, e.g. exciton complexes within p-shell or higher QD states. Nonetheless, it is more than sufficient to determine general energy structure of a single QD, where the energy spacing between the excited levels is usually significantly larger than the laser linewidth (e.g. 75 meV in the InGaAs/GaAs QDs presented above or see ref. [30]). The linewidth of the tunable laser in the Littman configuration can be further narrowed by the diffraction grating exchange, however it will influence the available tuning range and the output power. For the weaker QD signal detection, requiring long integration times, there can appear a problem with the sample drift (e.g. related to changes in helium pressure during the measurement at low temperatures), which can be mitigated by an automatic correction of the sample position between the consecutive measurements. There is also limited amount of wavelengths for which sharp longpass (or bandpass) filters are available. It can be overcome by replacing the longpass filter with a tunable bandpass filter formed by two diffraction gratings and a slit [31].

**Acknowledgements**

This work was supported by the National Science Centre (Poland) within the project 2014/15/D/ST3/00813. The authors thank S. Reitzenstein from Technische Universität Berlin, Germany, and M. Benyoucef from University of Kassel, Germany, for providing the quantum dot samples.

**Figure captions**

Fig. 1. Tunable laser excitation line with full width at half maximum values: (a) commercial external cavity laser in Littrow configuration, (b) self-made external cavity laser in the Littman configuration.

Fig. 2. Emission from single InAs/InP quantum dots with the nonresonant excitation by a standard diode laser (639 nm) and with excitation by an external cavity laser in Littrow configuration (1499 nm) before and after lasing filtering was applied. The edge of the longpass filter is marked by the dashed line.

Fig. 3. Scheme of the experimental setup: 1 – Littman configuration laser diode, 2 – lens, 3 – diffraction grating, 4 – adjustable mirror, 5 – mirror, 6 – commercial tunable laser in Littrow configuration, 7 – flip mirror, 8 – beamsplitter with a few percent of reflection, 9 – fixed-grating infrared spectrometer or the wavelength meter, 10 – shortpass filter, 11 – concave mirror, 12 – short-focal-length filtering monochromator, 13 – fixed-wavelength laser, 14 – motorized neutral density gradient filter, 15 – power meter head, 16 – beamsplitter cube, 17 – microscope objective, 18 – microscopy cryostat, 19 – infrared LED, 20 – iris diaphragm, 21 – infrared imaging camera, 22 – longpass filter, 23 – achromatic doublet lens, 24 – long-focal-length monochromator, 25 – InGaAs linear array detector.

Fig. 4. InGaAs/GaAs quantum dots: (a) single dot PLE map, (b) single dot PLE spectrum for 0.9258 eV emission line.

Fig. 5. InAs/InP quantum dots: (a) single dot PLE map, (b) single dot PLE spectrum for 0.7977 eV emission line.

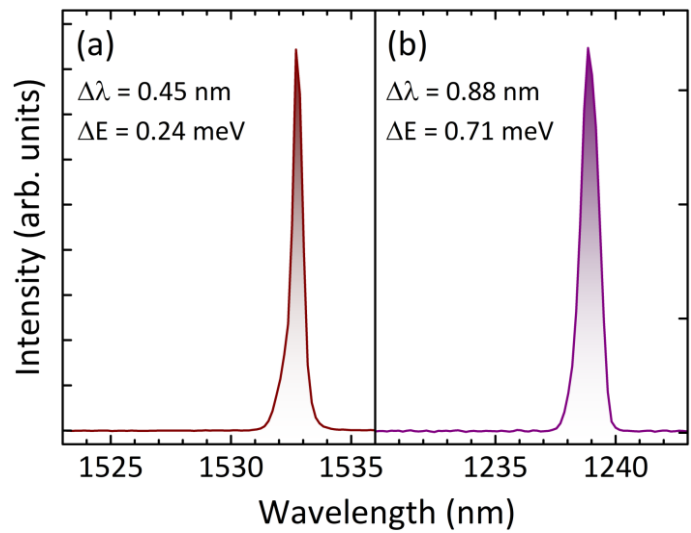

Fig. 1.

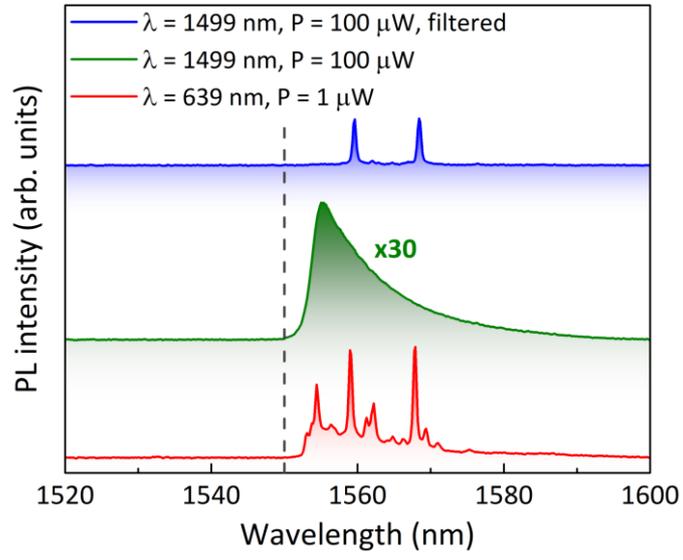

Fig. 2.

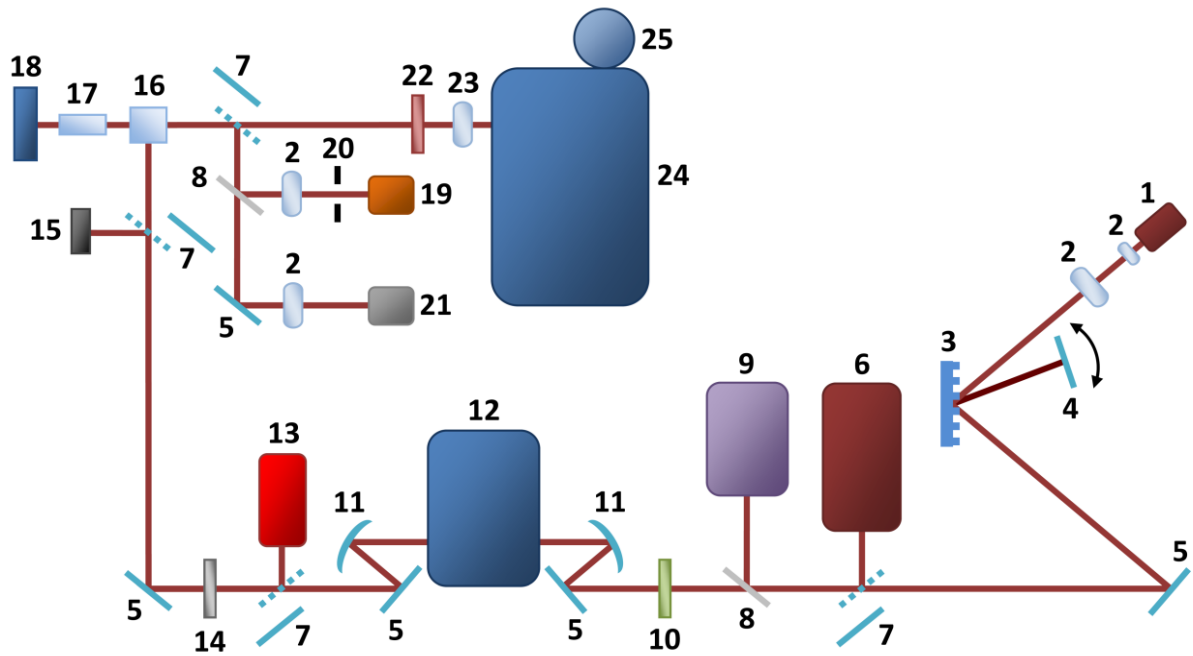

Fig. 3.

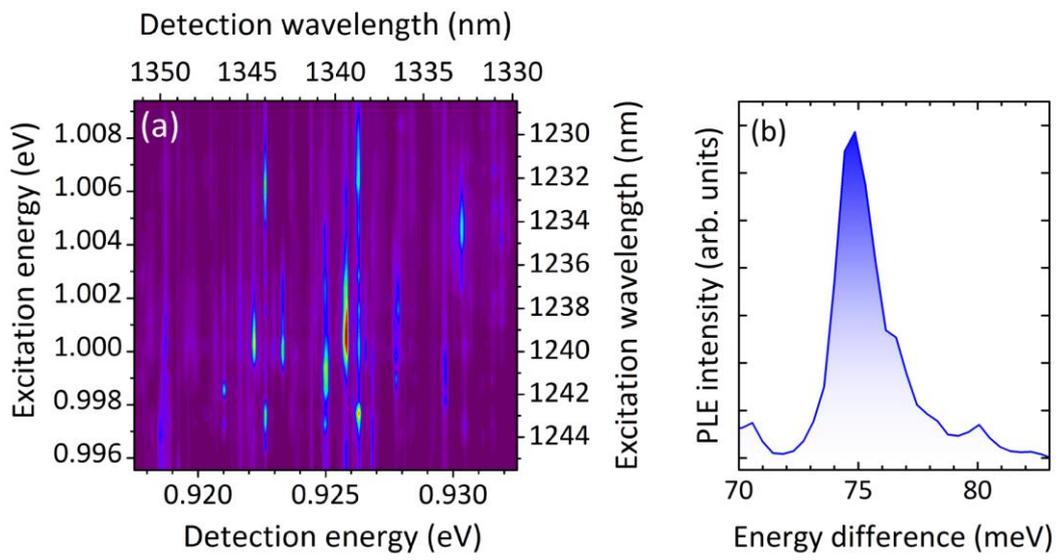

Fig. 4.

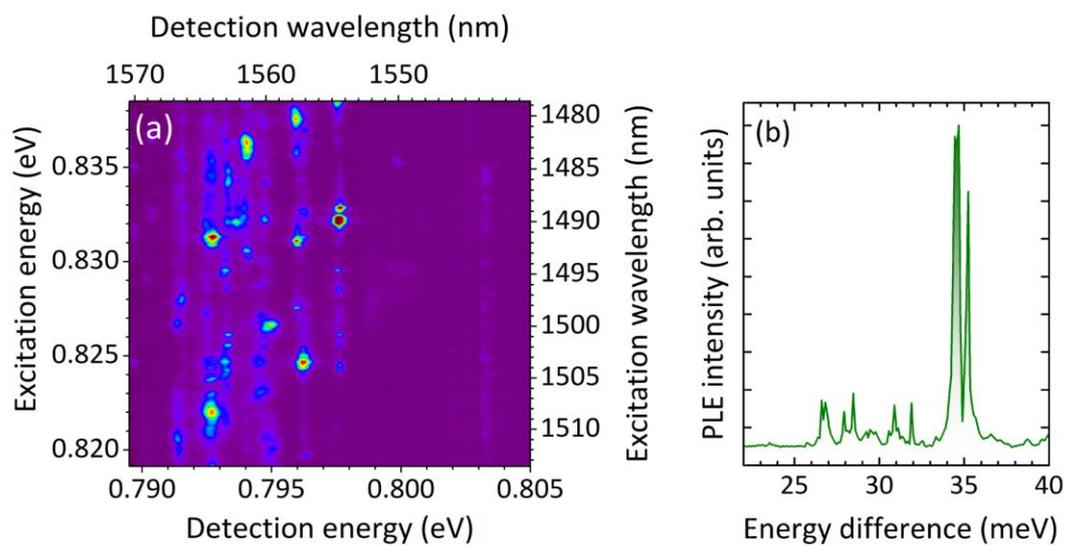

Fig. 5.